\documentclass[pra,10pt,aps,twocolumn,amsmath,amssymb,superscriptaddress,altaffilletter]{revtex4-1}
\usepackage{color}                                                     
\usepackage                             {lmodern}
\usepackage     [T1]                    {fontenc}
\usepackage                             {xcolor}
\usepackage                             {graphicx}
\usepackage     [english]               {babel}
\usepackage     [colorlinks,citecolor=blue,linkcolor=blue,urlcolor=blue,bookmarks=false,hypertexnames=true]           {hyperref}
\usepackage                             {dcolumn}
\usepackage                             {bm}
\usepackage                             {textcomp}
\usepackage                             {ulem}

\newcommand{\gm}[1]{{\color{black}{{#1}}}}

\begin{document}

\title{Fluorescence calorimetry of an ion crystal}

\author{Marvin Gajewski}
\email{marvin.gajewski@physik.uni-saarland.de}
\affiliation{Theoretische Physik, Universit\"at des Saarlandes, 66123 Saarbrücken, Germany}
\author{Wenbing Li}
\email{wenbingli@uni-mainz.de}
\affiliation{QUANTUM, Institut f\"ur Physik, Johannes Gutenberg-Universit\"at Mainz, 55099 Mainz, Germany}
\affiliation{Helmholtz-Institut Mainz, 55099 Mainz, Germany}
\author{Sebastian Wolf}
\affiliation{QUANTUM, Institut f\"ur Physik, Johannes Gutenberg-Universit\"at Mainz, 55099 Mainz, Germany}
\author{Walter Hahn}
\affiliation{Helmholtz-Institut Mainz, 55099 Mainz, Germany}
\affiliation{Institute for Quantum Optics and Quantum Information, Austrian Academy of Sciences, 6020 Innsbruck, Austria}
\author{Christoph\nolinebreak\ E.\nolinebreak\  D\"ullmann}
\affiliation{Helmholtz-Institut Mainz, 55099 Mainz, Germany}
\affiliation{PRISMA Cluster of Excellence, Johannes Gutenberg University Mainz, 55099 Mainz, Germany}
\affiliation{Department Chemie - Standort TRIGA, Johannes Gutenberg-Universität Mainz, 55099 Mainz, Germany}
\affiliation{GSI Helmholtzzentrum f\"ur Schwerionenforschung GmbH, 64291 Darmstadt, Germany}
\author{Dmitry Budker}
\affiliation{QUANTUM, Institut f\"ur Physik, Johannes Gutenberg-Universit\"at Mainz, 55099 Mainz, Germany}
\affiliation{Helmholtz-Institut Mainz, 55099 Mainz, Germany}
\affiliation{PRISMA Cluster of Excellence, Johannes Gutenberg University Mainz, 55099 Mainz, Germany}
\affiliation{Department of Physics, University of California, Berkeley, California 94720-7300, USA}
\author{Giovanna Morigi}
\affiliation{Theoretische Physik, Universit\"at des Saarlandes, 66123 Saarbrücken, Germany}
\author{Ferdinand Schmidt-Kaler}
\affiliation{QUANTUM, Institut f\"ur Physik, Johannes Gutenberg-Universit\"at Mainz, 55099 Mainz, Germany}
\affiliation{Helmholtz-Institut Mainz, 55099 Mainz, Germany}
\affiliation{PRISMA Cluster of Excellence, Johannes Gutenberg University Mainz, 55099 Mainz, Germany}

\begin{abstract}
Motivated by the challenge of identifying intruder ions in a cold ion crystal, we investigate calorimetry from emitted fluorescence light. Under continuous Doppler cooling, the ion crystal reaches a temperature equilibrium with a fixed level of fluorescence intensity and any change in the motional energy of the crystal results in a modification of this intensity. We theoretically determine the fluorescence rate of an ion crystal as a function of the temperature, assuming that laser light is scattered along a two-level electronic transition, which couples to the crystal's vibrations via the mechanical effects of light. We analyze how the heat dissipated by collisions of an incoming intruder ion alters the scattering rate. We argue that an energy change by an incoming $^{229}$Th$^{10+}$ ion can be unambiguously detected within 100 $\mu$s via illuminating \gm{a fraction} of a 10$^{3}$ ion crystal. This method enables applications including capture and spectroscopy of charged states of thorium isotopes and investigation of highly charged ions. 
\end{abstract}

\maketitle

\section{Introduction}\label{Intro}

Cold single trapped ions and ion crystals are well established workhorses in quantum optics, precision spectroscopy and quantum information. Ordered structures emerge from the combination of trapping forces and Coulomb repulsion at laser-cooling temperatures and realize Wigner crystals in the laboratory \,\cite{DubinONeilRMP:1999}. Due to the Coulomb interactions and the rarefied charge distribution, the thermodynamic properties are expected to exhibit peculiar features \,\cite{MorigiFishman:2004,BollingerDubinScience}. Due to the isolation from the environment, moreover,
measurements of temperatures, and in general, of the thermodynamic functions cannot be implemented as in normal solids.
This question acquires particular relevance for calorimetric studies. 

Calorimetry, namely, the determination of a transient change in temperature, might feature various application cases including the detection of sudden heating events which may be caused by collisions or by the injection of an ion or an ion bunch from an external source, coined here intruder ions. 
Apart from such applications, calorimetry of ion crystals may allow investigating the distribution and thermalization of impact energy over several motional degrees of freedom. 

Similar studies have been performed for a single trapped ion. In the so-called Doppler-recooling technique, time-dependent fluorescence emitted by a single trapped ion is recorded for characterizing the heating in the trap during a time interval when cooling light is turned off~\cite{wesenberg}. In a similar fashion, one may sense large coherent motional excitation after rapid ion shuttles via measuring the Doppler-recooling time \cite{Huber2008}. Recently, the fluorescence intensity emitted by a laser cooled crystal was numerically simulated for revealing information about intruding ions which heat up the crystal\,\cite{Champenois_MassSpectroscopy}. 

\begin{figure}[b]
 \centering
 \includegraphics[width=1.0\columnwidth]{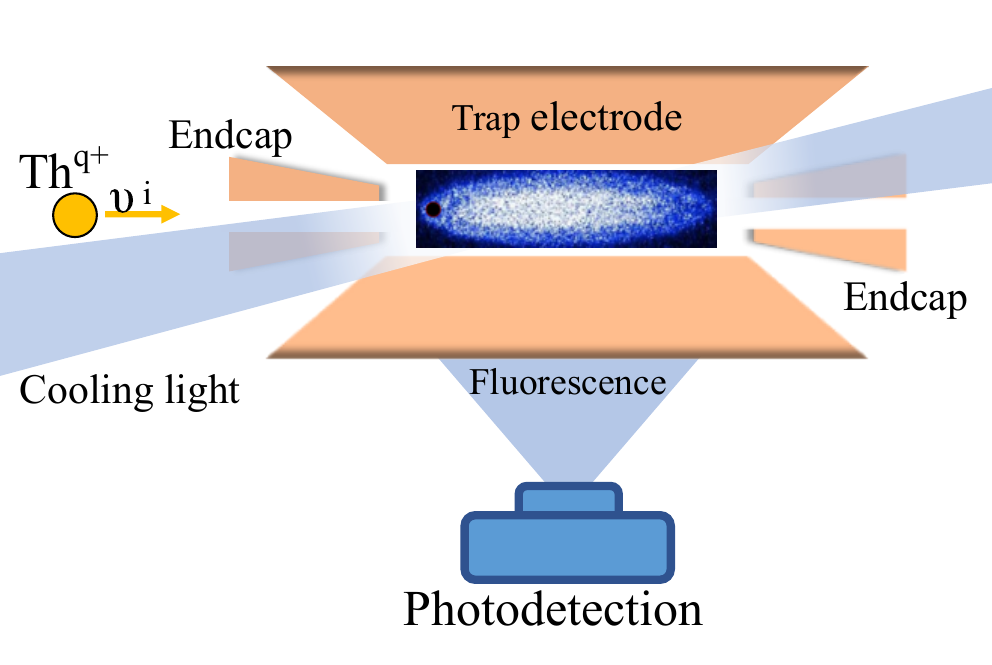} 
 \caption{Schematic of the setting considered. A thorium ion of charge state $q$ and velocity $v_\text{i}$ \gm{enters the Paul trap, containing a laser-cooled Ca$^+$ ion crystal. The ion enters from the left-hand side, by passing through the entrance endcap; it is repelled by the exit endcap potential (at the right-hand side)} and decelerates. \gm{We employ a beam near 397 nm which is directed along the axis such that the effects of micromotion are mitigated in the spectrum. An image of the laser-induced fluorescence for a large ion crystal is shown, where the black circle in the left-hand part indicates one successfully trapped thorium ion. The crystal heats up by impact of the intruder ion,} which leads to a change of intensity of the laser-induced fluorescence.}
\label{fig_sk}
\end{figure}

In this work we develop a theoretical model that allows us to systematically connect the mean vibrational energy of the crystal with the intensity of the scattered light. By means of our theoretical model we identify the optimal working conditions where the resonance fluorescence allows for a calorimetric measurement. Our starting point is the master equation for the internal and external degrees of freedom of $N$ ions forming a three dimensional ion crystal (Sec. \ref{sec_fluor}). In the regime in which the motion can be described by semiclassical equations\,\cite{stenholm1986semiclassical,Morigi_2001}, we  calculate the fluorescence emitted as a function of the temperature (Sec. \ref{sec_fluor_semi}). We study the dependence on laser detuning and laser power and identify optimal parameters for thermometry. We then apply this knowledge to reveal the injected energy from the change in the fluorescence  (Sec. \ref{sec_intrd_change}). As a specific application of fluorescence calorimetry of an ion crystal, we consider nondestructive detection of a single thorium ion stochastically entering (intruding) into an ion crystal of approximately $10^3$ $^{40}$Ca$^+$ ions. This is relevant for the trapped and sympathetically cooled thorium ions in calcium crystals (TACTICa) experiment (Fig. \ref{fig_sk}) \cite{Berning:2019,Stopp:2019,Hass:2020}, aiming at high-precision laser spectroscopic studies of thorium isotopes including the exotic low-energy nuclear isomer in $^{229}$Th \cite{wense2016,thielking2018}. This unique system is of interest for the study of effects unexplained by the standard model of particle physics \cite{flambaum2006,Peik2021nuclear} and for nuclear physics \cite{Beeks2021}.

While $^{229}$Th is a long-lived isotope and is available in macroscopic quantities, $^{229m}$Th is not \cite{Seiferle2017}; it is most readily available as the $\alpha$-decay daughter of $^{233}$U\,\cite{Raphael2020}. This isotope's $\alpha$ decay populates the isomeric state in about 2$\%$ of all cases. The decay produces ions with 84\,keV kinetic energy\,\cite{Hass:2020}, which needs to be removed for the successful trapping of the ion. In the TACTICa setup, this is achieved by putting the $^{233}$U recoil ion source at a potential of about --84/$q$\,keV, against which the daughter ions emitted in charge state $q$+ run up to arrive at the trap at near rest. Successful trapping relies on the ion entering the trap \gm{by the entrance endcap, (left-hand side in Fig.~\ref{fig_sk})} and being reflected by the \gm{repulsive potential generated by the} exit endcap. \gm{Finally on the way back, before the ion round trip time has elapsed, now} the entrance side is closed \gm{by applying a high repulsive potential}. Taking a realistic case example of thorium ions with charge states $q=1,\ldots,10$, we determine the dissipated energy of the intruding ion in an ion crystal of calcium (Sec. \ref{sec_intrd_energy}). We show that ion crystal fluorescence calorimetry will provide a non-destructive signal indicating intrusion of a $^{229}$Th ion, which can be used for triggering electrostatic closure of the trap. 

This paper is organised as follows. In Sec.\ \ref{sec_fluor} the theoretical model is derived. It is then applied in Sec.\ \ref{sec_intrd} to the detection scenario of an intruder ion. A summary and outlook to future work are given in Sec.\ \ref{sec_con}. Appendices \ref{app_cl}-\ref{app_ld} contain details of the calculations and of the parameters. 

\section{Measuring the temperature of large crystals with light} \label{sec_fluor}

In this section we derive the dependence of the fluorescence intensity on the thermal energy of an ion crystal.
This could in principle allow one to measure the temperature of an ion crystal. Moreover, it also can be applied to measure, e.g., the energy released by interaction with an intruder ion. For the purpose of developing a theory for performing thermometry using light, in this section we discuss a simplified model, which can be systematically derived from the full quantum-mechanical master equation of the ion crystal interacting with light. We argue that, despite the large number of degrees of freedom, a model based on only a few degrees of freedom allows one to describe most experimental scenarios of Doppler-cooled ion crystals.

\subsection{Model and basic assumptions} \label{sec_fluor_model}

We first start from the full quantum-mechanical model. We consider the motion of the ion crystal composed of $N$ ions,
which here can range from one to thousands \cite{DubinONeilRMP:1999}. Letting  ${\bf p_j}$ and ${\bf r_j}$ be the canonically conjugated momentum and position of ion $j$, respectively, and $M$ be the ion mass [without loss of generality we assume the ions to have equal mass, see Sec.\,\ref{sec_fluo_disc}], the mechanical energy is described by the Hamiltonian
\begin{equation}
\label{H:mec}
H_{\rm crystal}=\sum \frac{{\bf p_j}^2}{2M}+V_{\rm pot}({\bf r_1},\ldots,{\bf r_N})\,,   
\end{equation}
where the potential energy $V_{\rm pot}$ includes the repulsive Coulomb interaction and the trap potential. For sufficiently cold ion crystals, the Hamiltonian \eqref{H:mec} can be reduced to describe the energy of a harmonic crystal, whose eigenmodes are the harmonic collective vibrations of the ions about the equilibrium positions of potential $V_{\rm pot}$. 

The motion couples with light via the mechanical effects of atom-photon interactions. Laser light drives an optical transition of the ions, with ground state $|g\rangle$ and excited state $|e\rangle$. Let ${\bf d}$ be the dipole moment and $\gamma$ the spectral linewidth of the transition. We define by $\Delta=\omega_0-\omega_L$ the detuning of the laser frequency $\omega_L$ (for Doppler cooling $\omega_L<\omega_0$ \cite{wineland,stenholm1986semiclassical,Eschner:03}). The total Hamiltonian reads $H=H_{\rm crystal}+H_{\rm ions}+H_{\rm AL}$, where $H_{\rm ions}=\sum_j\hbar\Delta |e\rangle_j\langle e|$ and the laser-dipole interactions, in the electric-dipole approximation, take the form
\begin{equation}
H_{\rm AL}=\sum_j\frac{\hbar\Omega({\bf r_j)}}{2}[\sigma_j^\dagger\exp(-i\mathbf{k\cdot r_j})+\sigma_j\exp(i\mathbf{k\cdot r_j})]\,.
\end{equation}
In this description the laser is a classical traveling wave with wave vector $\mathbf{k}$ and position-dependent Rabi frequency $\Omega({\bf r_j})=-{\bf d}\cdot{\bf E_L ({\bf r_j})}/\hbar$, where ${\bf E_L}({\bf r_j})$ is the slowly-varying amplitude of the laser electric field. Here,  $\sigma_j=|g\rangle_j\langle e|$  is the lowering operator for the internal electronic transition of ion $j$ and $\sigma_j^\dagger$ is its adjoint. 

The dynamics of internal and external degrees of freedom of the crystal results from the interplay between the coherent dynamics, governed by the Hamiltonian $H$, and the incoherent one due to the spontaneous decay. It is described by the master equation for the density matrix $\varrho$:
\begin{equation} \label{eqn_mastereq}
	\frac{\partial\varrho}{\partial t}=-\frac{i}{\hbar}\left[ H,\varrho\right]+L(\varrho)\,,
\end{equation}
where $L\varrho=\gamma/2\sum_j[2\sigma_j\tilde{\varrho}\sigma_j^\dagger-\sigma_j^\dagger\sigma_j\varrho-\varrho\sigma_j^\dagger\sigma_j]$ is the superoperator for the spontaneous decay, which includes the diffusion due to the stochastic mechanical forces associated with the emission of a photon with wave vector ${\bf k'}$, namely,  $\tilde{\varrho}\propto\sum_j\int {\rm d}{\bf k'} f({\bf k'}) e^{i\mathbf{k'\cdot r_j}}\rho e^{-i\mathbf{k\cdot r_j}}$. The probability density $f({\bf k'})$ weights the process and depends on the dipole's polarization. We here discard collective effects in the spontaneous emission as the interparticle distances between ions are of the order of several optical wavelengths.

In this work we determine the rate of photon emission
\begin{equation}
    S=\gamma\sum_j{\rm Tr}\{|e\rangle_j\langle e|\varrho(t)\}
\end{equation}
and its dependence on the temperature of the crystal when the crystal is in a thermal-equilibrium state.

\subsection{Experimental parameters} \label{sec_fluor_param}

Modeling the dynamics of resonance fluorescence requires taking into account the coupled equations of internal and external degrees of freedom of the ions forming the crystal. For a large crystal this is a formidable task. This complexity can be significantly reduced in the case of Doppler cooling, and more generally, when the characteristic frequency scales are orders of magnitude larger than the recoil frequency $\omega_R=\hbar k^2/2M$. These assumptions are valid for the parameters of typical experimental ion-trap settings. For example, in the TACTICa experiment\,\cite{Berning:2019,Stopp:2019,Hass:2020}
the crystals are composed of $^{40}$Ca$^+$ trapped ions. 
The secular trap frequencies $\omega_\text{x,y,z}/2\pi$ are lower than \gm{1 MHz (see Appendix \ref{app_osc})} and determine the smallest characteristic frequency of the motion
(provided the system is sufficiently far away from structural instabilities\,\cite{Birkl_1994,DubinONeilRMP:1999,Fishman:2008}). 
The natural linewidth of the S$_{1/2}$ to P$_{1/2}$ transition in $^{40}$Ca$^+$ near 397\,nm is $\gamma/2\pi$ = 21.6\,MHz which corresponds to a lifetime of  6.904(26)\,ns\,\cite{hettrich}. The recoil frequency is $\omega_R=2\pi\times31.5$\,kHz, thus it is one order of magnitude smaller than the characteristic frequencies determining the crystal dynamics. The time scale $\tau_\text{cool}$ of the mechanical effects of light is of the order of \gm{200\,$\mu$s} for a single ion driven below saturation (see Appendix \ref{app_ld}). When a single ion sympathetically cools a crystal composed of $N$ ions, a very rough estimate gives a cooling time that scales as $\tau_\text{cool,N}\sim N\tau_\text{cool}$. The same rough estimate allows us to give a figure of merit for the cooling timescale, if $M$ ions sympathetically cool the crystal and the ions are driven below saturation. In this case the cooling time decreases as $\tau_\text{cool,N}/M$ \cite{Morigi_2001}. As long as the cooling time scale is significantly longer than the lifetime of the excited state, we can apply the time scale separation ansatz at the basis of our treatment.   

\subsection{Time-scale separation} \label{sec_fluor_time}

This parameter regime allows us to make some simplifications based on the following observations. In the first place, 
the internal degrees of freedom evolve over a faster time scale 
than the timescale $\tau_\text{cool}$ of the external degrees of freedom.  In this limit internal and external degrees of freedom are not entangled to a good approximation and the density matrix of the ions is separable $\varrho\approx\rho\otimes \mu$, with $\rho$ and $\mu$ the density matrices for the internal and  the external degrees of freedom, respectively. We can hence solve the equations of motion for $\rho$ over a time scale such the internal degrees of freedom relax to the steady state \gm{$\rho^\text{(SS)}$}, while the external degrees of freedom do not have time to evolve. Let $\{|{\cal E},\ell\rangle\}$ be a basis of eigenstates of $H_{\rm crystal}$ with mechanical energy $\cal E$, where $\ell$ is a set of quantum numbers which uniquely determines the quantum state. We now write the optical Bloch equations for $\rho$ assuming that only one ion (say, ion $j$) of the crystal is coupled to the laser light and that the spatial displacement of this ion couples to all normal modes. We use the notation $$\rho_{pq}(\mathcal E,\mathcal E')\equiv\sum_\ell\langle p_j,\mathcal E,\ell|\varrho|q_j,\mathcal E',\ell\rangle\,,$$ with $p,q=e,g$, and obtain the corresponding equations of motion:
\begin{align} \label{eqn_bloch}
    \frac{d}{dt} \rho_{ee}(\mathcal E,\mathcal E') &= -\gamma \rho_{ee}(\mathcal E,\mathcal E') \\
    &+ \frac{i\Omega}{2}\sum_{\mathcal E_1}\left[T(\mathcal E,\mathcal E_1)\rho_{ge}(\mathcal E_1,\mathcal E') - \text{c.c.}\right], \nonumber \\
    \frac{d}{dt}\rho_{eg}(\mathcal E,\mathcal E') &= -\left(\frac{\gamma}{2} -i\Delta -i\frac{\mathcal E-\mathcal E'}{\hbar} \right) \rho_{eg}(\mathcal E,\mathcal E') \\
    &+ \frac{i\Omega}{2}\sum_{\mathcal E_1}T(\mathcal E,\mathcal E_1)\left[\rho_{gg}(\mathcal E_1,\mathcal E') - \rho_{ee}(\mathcal E_1,\mathcal E') \right]\,,\nonumber 
\end{align}
where $\rho_{ge}(\mathcal E,\mathcal E')=\rho_{eg}(\mathcal E',\mathcal E)$ and the coefficient 
\begin{equation}
T(\mathcal E,\mathcal E')=\langle \mathcal E|  \exp(i{\bf k}\cdot {\bf r_j})|\mathcal E'\rangle\  
\end{equation}
is the probability amplitude of a transition from the mechanical state $|\mathcal E'\rangle$ to the mechanical state $|\mathcal E\rangle$ by absorption of a photon. Moreover, under the time-scale-separation ansatz, we neglect the effect of the recoil of spontaneous emission on the motion, which gives rise to diffusion \cite{stenholm1986semiclassical,Morigi_2001}.
Therefore, the number of coupled equations to solve is finite. Since $\gamma\tau_{\rm cool}\gg 1$, we can find a time scale $\Delta t$ such that $\Delta t\gg 1/\gamma$ and $\Delta t\ll\tau_{\rm cool}$. Hence, we can neglect a change of the motional state over the time scale $\Delta t$ and determine the internal state by setting 
$\dot\rho_{ee}^\text{(SS)}(\mathcal E,\mathcal E')=\dot\rho_{eg}^\text{(SS)}(\mathcal E,\mathcal E')=0$. The resonance fluorescence signal over the time scale $\Delta t$ is 
\begin{equation}
    S=\gamma\sum_{\mathcal E}\rho_{ee}^\text{(SS)}(\mathcal E,\mathcal E)p(\mathcal E)\,,
\end{equation}
where 
$p(\mathcal E)$ is the probability that the crystal has mechanical energy $\mathcal E$.

\subsection{Semiclassical limit} \label{sec_fluor_semi}

We now perform a further simplification that reduces the number of equations to solve to two. This simplification relies on (i) a coarse graining in energy, which can be made in the limit where the ion-crystal temperature $T$ is $k_BT\gg \hbar\omega_R$ and (ii) the weak-binding regime, when the spectral linewidth $\gamma\gg {\rm max}_{n}\omega_n$, where $\omega_n$ are the vibrational frequencies of the crystal and $n$ labels the mode \cite{stenholm1986semiclassical,Morigi_2001} (see Appendix \ref{app_osc}). The equations are determined for an average energy $\mathcal E$ over the coarse-grained energy scale $\delta E$ such that $k_BT\gg \delta E\gg\hbar\omega_R$ \cite{Morigi_2001}. At the leading order in the expansion in the small parameter $\hbar\omega_R/k_BT$, the solution of the corresponding optical Bloch equations reads 
\begin{align}
   \rho_\text{ee}^\text{(SS)}(E) = \frac{\Omega^2/4}{\Delta_\text{eff}(E)^2 + \gamma^2/4 + \Omega^2/2}\,, \label{eq:obe_solution}
\end{align}  
where
\begin{align}
    \Delta_\text{eff}(E) = (E_e-E_g)/\hbar \label{eq:Delta_eff}
\end{align}
and is the frequency gap between the ground and excited states coupled by the absorption of a laser photon. In the weak-binding limit the ground and excited state energies can be written as
\begin{align*}
    E_g =& \frac{\textbf{p}^2}{2m} + V(\textbf{r}) , \\ 
    E_e =& \frac{(\textbf{p}-\hbar\textbf{k})^2}{2m} + V(\textbf{r}) + \hbar\Delta\,.
\end{align*}
We emphasize that these expressions for the energy are valid when one can neglect the spatial displacement of the ions during a photon-scattering event. Let us summarize when this is a valid assumption. For this purpose, we consider a single ion in one dimension, and remark that the argument can be straightforwardly extended to the case when the motion is in three dimensions and the scattering ion is embedded in a crystalline structure. For a single ion in a trap at frequency $\omega$, the spatial displacement after a photon-scattering event is given by $\Delta x\sim p/M\gamma$, where $p\sim\sqrt{k_BTM}$. This is much smaller than the oscillation amplitude $\mathcal A_x\sim \sqrt{k_BT/M\omega^2}$, when $\omega\ll\gamma$ \cite{stenholm1986semiclassical}. 
As a result, in this regime the population of the excited state solely depends on the ion-momentum distribution at leading order in an expansion in $\omega/\gamma$ (for an ion crystal, the small parameter is $\omega_n/\gamma$).

\begin{figure*}[t!] 
	\centering
    \includegraphics[width=2.05\columnwidth]{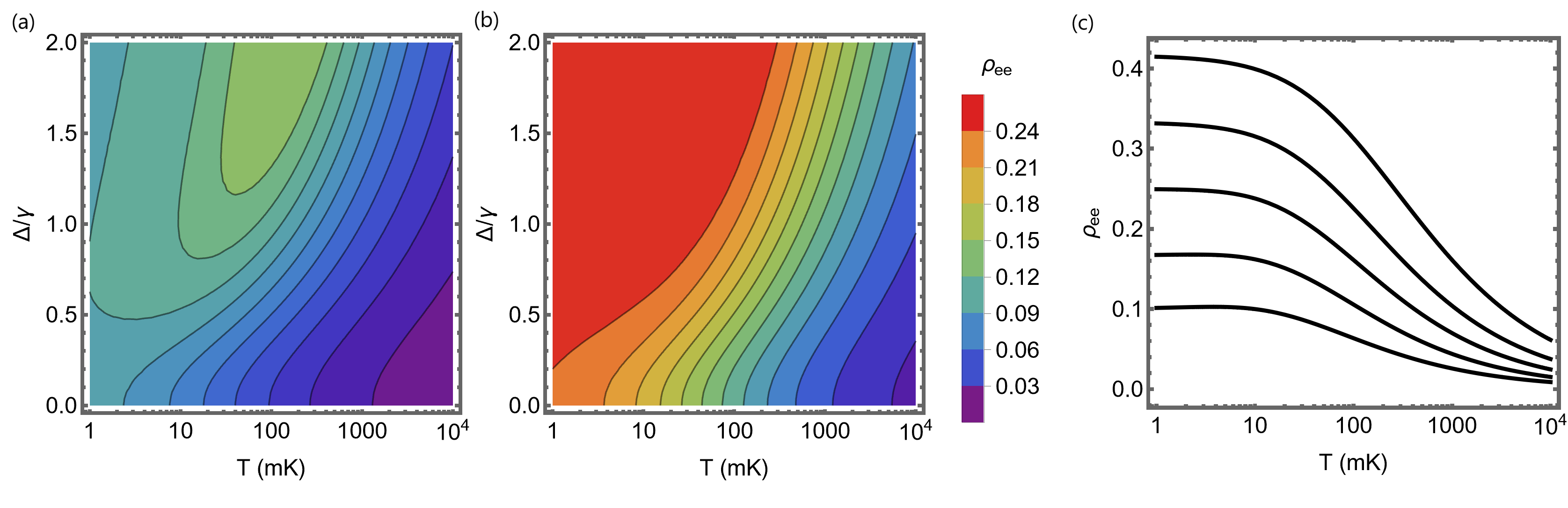}
	\caption{Fluorescence $S=\gamma\rho_{ee}$ of an ion crystal {when a single ion of the crystal is illuminated}. The plot displays $\rho_{ee}$, the average probability to be in the excited state {of the illuminated ion} as a function of the temperature. Here $\rho_{ee}(T)$ is plotted as a function of the temperature $T$ and of the detuning $\Delta$ (in units of $\gamma$) for a fixed saturation parameter (a) $s=0.25$ and (b) $s=1$ [note that, as we vary $\Delta$, we accordingly vary $\Omega$ in order to keep $s$ constant; see Eq. \eqref{eq_sat}]. (c) Plot of $\rho_{ee}(T)$ as a function of $T$ for $\Delta = \gamma/2$ and for the saturation parameter $s=0.25, 0.5, 1, 2, 5$ (lowest to highest curve). {The behavior for $s\ll 1$ applies also to the case in which $N$ ions of the crystal are illuminated below saturation (see Sec. \ref{sec_fluo_disc}).} }
	\label{fig_sat}
\end{figure*}

In the following we assume that the ion phase-space function is a Boltzmann distribution, $\exp(-\beta H_{\rm crystal})$, where $\beta^{-1}=k_B T$ with $T$ the temperature.
Then, the total excited state population is given by the expression \cite{wineland}
\begin{align}
    \rho_\text{ee}(T) &= \int \rho_\text{ee}^\text{(SS)}(\bar E) \frac{e^{-\beta H_{\rm crystal}}}{\mathcal Z} d^{3N}\textbf{p}d^{3N}\textbf{r}\,, \label{eq:rhoee_general}
\end{align}
where $\mathcal Z$ is the partition function. We insert Eq. \eqref{eq:obe_solution} into Eq. \eqref{eq:rhoee_general}, discard higher orders in $\omega_R$, consistent with the validity of the semiclassical expansion, and rescale the variables using the saturation parameter $s$, which is defined as
\begin{equation}
s=\frac{\Omega^2/2}{\gamma^2/4+{\Delta'}^2}\,,   \label{eq_sat}
\end{equation}
with $\Delta'=\Delta+\omega_R$. Since the population of the excited state of ion $j$ solely depends on the momentum of the same ion, all other degrees of freedom can be integrated out and Eq.\,\eqref{eq:rhoee_general} is then brought to the form
 \begin{align}
\rho_\text{ee}(T) &= \sqrt{\frac{1}{\pi}} \int_{-\infty}^\infty \frac{s e^{-p_z^2}}{1 + 2s+\frac{(\zeta p_z)^2-2\Delta' \zeta p_z}{\gamma^2/4+{\Delta'}^2}} dp_z\,, \label{eq:rhoee}
\end{align}
\noindent where $\zeta=2\sqrt{\omega_R/\hbar\beta}$. Without loss of generality, in the above equation we assumed that laser light propagates along the $z$ direction. From the expression \eqref{eq:rhoee} we can directly find the resonance fluorescence: $S(T)=\gamma\rho_\text{ee}(T)$. Figures \ref{fig_sat}(a) and \ref{fig_sat}(b) display Eq. \eqref{eq:rhoee} for two values of the saturation parameter $s$ and as a function of the temperature $T$ and of the detuning $\Delta$. At low saturation [Fig. \ref{fig_sat}(a)] the resonance fluorescence signal reaches a maximum as a function of $T$. This maximum depends on the detuning and moves towards higher values of $T$ with increasing $\Delta$. On the contrary, at saturation [Fig.~\ref{fig_sat}(b)] the resonance fluorescence signal is a monotonically decreasing function of $T$. The temperature dependence of the resonance fluorescence at fixed $\Delta$ and different saturation parameters $s$ is illustrated in Fig. \ref{fig_sat}(c). We emphasize that the excited state population of an ion embedded in a crystal [Eq. \eqref{eq:rhoee_general}] depends on the crystal's motional energy in the same way as when the crystal consists of a single trapped ion (see also Ref. \cite{Biesheuvel2016}). We have shown that this holds when we can assume that the time scale in which a photon is scattered is much faster than the time scale over which a signal propagates across the crystal. The light scattered by the ion allows one to access the crystal's temperature assuming the system is at equilibrium. This assumption requires, among other necessities, that the micromotion can be neglected and the trapping potential can be faithfully described by the secular approximation. We refer the reader to Refs.\ \cite{Champenois_MassSpectroscopy, micromotion} for a detailed characterization of the effects of micromotion.

\subsection{Discussion} \label{sec_fluo_disc}

We now comment on the validity of our model and on the limits of its applicability. The model has been derived under the assumption that the relevant electronic degrees of freedom are two. In this limit we could find a closed expression for the fluorescence as a function of the temperature. This strictly holds when a single ion of the crystal is illuminated. Nevertheless, from this result one can extract the resonance fluorescence for the case when the ion crystal is illuminated, as long as the laser drives the ions below saturation. In that case, given the saturation parameter $s$ for a single ion, our model requires that $Ns\ll 1$. \footnote{When the laser drives the ions well below saturation, in fact, one can reduce the problem to a two-level system, consisting of the state $|G\rangle$, where all the ions are in the ground state, and the Dicke state $|E\rangle$, which is coupled to $|G\rangle$ by absorption of a laser photon and contains one (delocalized) electronic excitation. Then the resonance fluorescence over the full space is determined by the population of $|E\rangle$ as given by Eq. \eqref{eq:obe_solution} after replacing $\Omega\to \sqrt{N}\Omega$. We remark that this is strictly correct provided the laser parameters are homogeneous across the crystal and that $Ns\ll1 $, with $s$ given in Eq. \eqref{eq_sat}, so that, on average, only one ion can be excited at a time \cite{Morigi_2001,Morigi2003}. The ions can be considered as independent scatterers also at moderately larger intensities, as long as the temperature is sufficiently large.} The extension to the case in which the laser parameters are not homogeneous can be performed along the lines of the study in Ref. \cite{Schmit:2021}: In this case $\rho_{ee}$ [Eq. \eqref{eq:obe_solution}] becomes a function of the position inside the crystal and the resonance fluorescence is found by averaging $\rho_{ee}$ also over the spatial distribution of the laser parameter across the crystal. We remark that our model does not capture the general case, when the internal state of the emitters can be multiply excited at the same time and multiple absorptions of laser photons create correlations between internal and external degrees of freedom. In this regime, in fact, the set of optical Bloch equations that will be solved is of the order of $2^N$ and we expect that the dependence on the mean energy and on the saturation parameter takes a different functional form.  

Our model neglects the ions' displacement during the scattering process and thus requires that the lifetime of the excited state is much smaller than the oscillation frequency, more specifically, $\omega_n \ll \gamma$. This is the so-called weak-binding limit \cite{wineland}, which is consistent with a semiclassical description provided the recoil frequency is smaller than the linewidth\,\cite{stenholm1986semiclassical,Morigi_2001}. 
In this limit, the result does not depend on the mode spectra, and thus not on the ion masses within the crystal, as long as the modes are at the same temperature and the scattering ions are of the same species.

We finally comment on whether the approximations adopted here describe the setup of TACTICa. The semiclassical limit requires temperatures $T_\text{min} = \hbar\omega_R/k_B$. For the parameters of the TACTICa experiment, $T_\text{min} \sim 1 \mu$K, which is three orders of magnitude smaller than the Doppler-cooling temperature. The weak-binding regime requires that $\omega_n\ll\gamma$. According to our estimates, this is approximately fulfilled since $\max\omega_n \sim \gamma/3$.  

\section{Detection scenario of an intruder ion} \label{sec_intrd}

We consider the scenario of identifying a sudden addition of motional energy to the ion crystal experiencing Coulomb collisions of an intruder ion. While the ion crystal is kept under continuous Doppler cooling, this impact will change the fluorescence rate of the emitted light which is recorded (see Fig. \ref{fig_sk}). We discuss the example of a $^{40}$Ca$^+$ ion crystal under Doppler cooling on the S$_{1/2}$ to P$_{1/2}$ dipole-allowed transition near 397 nm and a single incoming $^{229}$Th$^{q+}$ ion as the intruder. In this regime the motion can be described by the semiclassical model (see Sec. \ref{sec_fluo_disc}). Let the thorium charge state $q$ range from +1 to about +10 elementary charges using the uranium recoil source\,\cite{Hass:2020}. The analysis of the detection scenario is based on three main steps: First, we estimate the energy increase of the crystal from the impact by the intruder, then we use the results of Sec. \ref{sec_fluor_semi} to calculate the corresponding change of fluorescence emission, and finally we discuss the feasibility of the proposed detection scheme taking into account the experimental parameters. Our results, as outlined below, confirm a recently published numerical simulation \cite{Champenois_MassSpectroscopy}. The analytical solution, moreover, may pave the way for an understanding of heat capacity in such model systems and intruder-induced heating effects.      

\subsection{Energy increase of the crystal due to an intruder ion} \label{sec_intrd_energy}

We model the deceleration of a Th$^{q+}$ ion in a Ca$^+$ ion crystal using \cite{POTH,ZWICKNAGEL}
\begin{equation} \label{eqn_dedx}
\frac{dE}{dx}\approx-\frac{q^2e^2}{4\pi\epsilon_0}\frac{\nu_\text{p}^2}{v^2(x)}\Lambda_\text{C},
\end{equation}
where $E$ is the kinetic energy of the Th$^{q+}$ ion at velocity $v(x)$, $x$ is the distance traveled in the ion crystal, $\nu_\text{p}$ is the plasma frequency of the ion crystal, and $\Lambda_\text{C}$ is a constant usually referred to as the Coulomb logarithm (see Appendix\,\ref{app_cl}). The plasma frequency $\nu_\text{p}$ describes the characteristic timescale for the response of the ion crystal to external perturbations and can be expressed as $\nu_\text{p}=\sqrt{e^2n_\text{d}/\epsilon_0M}$, where $n_\text{d}$ is the density of the ions of mass $M$ in the ion crystal (40 $u$ for Ca$^+$ ions). The expression \eqref{eqn_dedx} is based on linear response \cite{POTH,ZWICKNAGEL} and was shown to provide reliable estimates \footnote{The expression \eqref{eqn_dedx} remains valid until the velocity of the thorium ion becomes comparable to the mean thermal velocity of individual calcium ions, but this limitation is of minor importance here.} of the order of magnitude by means of numerical simulations \cite{BUSSMANN}.

The voltage on the entrance-endcap electrode, through which the intruder ion enters, is initially adjusted to just allow it entering and the potential on the opposite electrode is set for reflection. The goal is to detect the intruder ion and pulse the potential on the entrance-endcap electrode to close the trap before the intruder ion escapes again after its round-trip. Since the energy transfer becomes more efficient for intruders at lower energy [$dE/dx\propto1/v^2(x)$; cf. Eq. \eqref{eqn_dedx}], the maximal energy deposited by the thorium ion after a round-trip in the crystal is when it would completely stop after the round-trip, an optimal case on which we focus now.

To estimate the characteristic energy of the above process, we obtain from Eq. \eqref{eqn_dedx} $v(x)\approx(v_\text{i}^4-c_0x)^{1/4}$, where $v_\text{i}$ is the initial velocity of the $^{229}\text{Th}^{q+}$ ion, \mbox{$c_0\equiv q^2e^2\nu_\text{p}^2\Lambda_\text{C}/\pi\epsilon_0M_\text{Th}$}, and $M_\text{Th}=229$ u is the mass of $^{229}\text{Th}$. Assuming an ion trajectory on the trap axis, the thorium ion  is stopped when its initial velocity is \mbox{$v_{\text{i}2}\approx(2Lc_0)^{1/4}$}, where $L$ is the length of the crystal in axial direction. Substituting $\nu_\text{p}\approx 3$ MHz and $L\approx1$ cm for the TACTICa experiment, we obtain $v_{\text{i}2}\approx 2.5\cdot  10^2$ m/s for $q=1$, corresponding to an energy of $E_{\text{i}2}\approx75$ meV. For $q=10$, the energy increases to $E_{\text{i}2}\approx0.75\,$eV because $E_{\text{i}2}\propto q$. In the case of a uranium $\alpha$-recoil ion source, thorium ions are delivered at kinetic energies of 70--100 keV. For efficient Coulomb coupling with the Ca crystal, we plan to set the source at high attractive electrostatic potential for decelerating emitted Th ions \,\cite{Hass:2020}.

In the Coulomb crystal, ion motion is described by collective modes, such that the absorbed energy is quickly distributed, faster than the Doppler cooling dynamics. For the particular case of a single Th$^+$ ion stopped after traversing the ion crystal twice and the above parameters, we estimate a temperature increase $\delta T\sim 10$ mK, while for a Th$^{10+}$ ion, we obtain $\delta T\sim 100$ mK. 

\subsection{Change of fluorescence} \label{sec_intrd_change}
\begin{figure*}
	\centering
	\includegraphics[width=2.05\columnwidth]{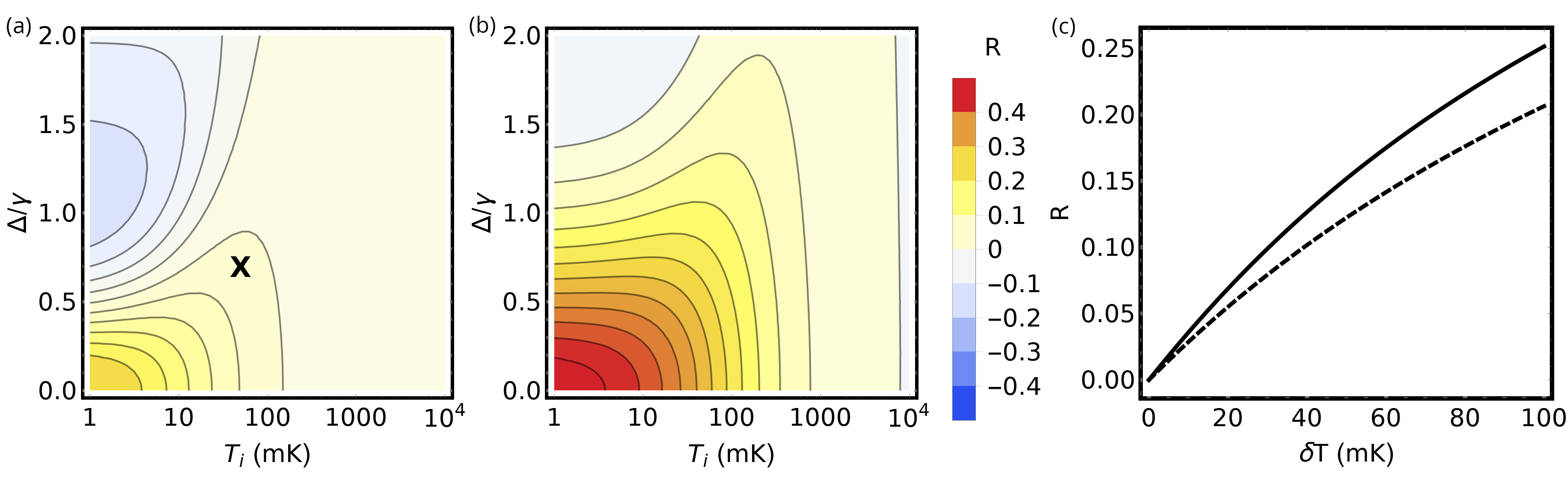} 
	\caption{Relative decrease of fluorescence intensity $R(T_i,\delta T)$ [cf. Eq. \eqref{eqn_r}] over an increase of temperature $\delta T$. Here $R(T_i,\delta T)$ is plotted as a function of initial temperature $T_i$ (in mK) and $\Delta$ (in units of $\gamma$) and for (a) $s=0.25$ and $\delta T = 10$ mK and (b) $s=1$ and $\delta T = 100$ mK. The scenario of a $^{229}$Th$^{\gm{q}+}$ intruder ion discussed in Sec. \ref{sec_intrd} corresponds to the black cross (for $q=1$). (c) Plot of $R(T_i,\delta T)$ as a function of $\delta T$ (in mK) for an initial temperature of $T_i = 50\,$mK, detuning $\Delta = 3\gamma/4 $, and $s=1$ (dashed line) and $s=0.25$ (solid line), which corresponds to $q\leq 10$ and the discussed example in the text. The curves have been evaluated for the case when a single ion of the crystal is illuminated. For $s\ll 1$ they also describe the regime in which all ions of the crystal are uniformly driven by the laser below saturation (see Sec. \ref{sec_fluo_disc}). }
	\label{fig_R}
\end{figure*}

We have investigated the dependence of fluorescence intensity on crystal temperature in Sec. \ref{sec_fluor_semi}. A convenient observable used to measure the change in fluorescence is the relative decrease, quantified by
\begin{equation} \label{eqn_r}
R(T_i,\delta T)\equiv\frac{S(T_i + \delta T)-S(T_i)}{S(T_i)},
\end{equation}
where $S(T) = \gamma \rho_{ee}(T)$ is the measured intensity of fluorescence at temperature $T$, and $T_i$ and $\delta T$ are the initial temperature of the ion crystal and the temperature shift, respectively. We illuminate $M$ ions of a $10^3$ ion crystal with a saturation per ion $s\ll 1$ such that $Ms\approx 0.5$. Taking into account the case example of a single Th$^{10+}$ ion intruding into a Coulomb crystal at temperature $T_i=50$ mK, with detuning $\Delta=15$ MHz, then the relative decrease of fluorescence is $R(T_i,100\text{ mK})\sim 0.25$ (see Fig. \ref{fig_R}). The change takes place over a time of about \gm{$100 \mu$s}, until the deposited energy is removed by Doppler cooling over a timescale of several milliseconds (see Sec. \ref{sec_fluor_param}). The relative decrease of fluorescence $R(T_i,10\text{ mK})=0.04$ is less significant for a singly charged Th intruder. 

\subsection{Efficiency of detection} \label{sec_intrd_eff}

The Ca crystal fluorescence is continuously observed by a photon counting unit to discriminate the intruder event of a single Th$^{10+}$ from detecting the relative change of fluorescence of $R(T_i,\delta T)$. We assume  a scattering rate $\gm{\gamma}\rho_{ee} $ per ion, and take into account an overall photon detection efficiency of approximately $5\%$, illuminating a fraction $M$ of a crystal composed of 10$^{3}$ ions with a relative fluorescence change of $R=0.25$. Employing Poissonian counting statistics, the tagging signal would exceed noise by approximately 1.1 standard deviations already within a counting time of 100 $\mu$s, which is clearly shorter compared to the Doppler-recooling time. In the case of a singly charged intruder ion, however, the relative change in fluorescence $R$ is 0.04, and the event detection with approximately 1.1 standard deviation would require a 5 ms counting time. Thus, for a singly charged intruder ion, the necessary detection time would be too long, as this time may possibly exceed the Doppler-recooling time. 
An increase in the number of illuminated ions would go along with a better optical signal. However, one has to account for the time scale of the recooling dynamics, which limits the maximum observation time.
Therefore, for an efficient detection of a singly charged intruder ion, we propose improving the detection optics e.g. using high numerical aperture objectives, and the quantum efficiency of the detector, furthermore reducing the initial cooling limit of the crystal to enlarge $R$.

Note that tagging of small bunches of Ar$^+$ ions at 4~keV has been demonstrated, based on their flyby through a narrow electrode structure and the detection of an induced voltage. For bunches of about 300 ions, the detection signal exceeds the amplifier noise level~\cite{RAECKE}. Thus, our method features an alternative for loading Paul traps deterministically at the single particle level with intruder ions, that might be stemming either from electron beam ion trap sources~\cite{MICKE18,ANDELKOVIC2015109} or from radioactive sources~\cite{Hass:2020} with a perspective of precision quantum logic spectroscopy using exotic matter~\cite{MICKE20}.

\section{Outlook} \label{sec_con}

We have developed an analytical model that allows us to determine the dependence of the resonance fluorescence of an ion crystal on its temperature and thus to access the heat capacity of the crystal. We apply this model to investigate intruder-induced heating, as it is detected from the modification of laser-induced fluorescence.  We envision direct applications for the capture of singly charged thorium ions and eventually also of more highly charged ions. 

Our model is based on the assumption that the signal is collected over a time that is smaller than the time scale of 
the mechanical effects of light.
Analysis of the timescales showed that during the recooling period a state-of-the-art trapped-ion setup will be able to collect a sufficiently large number of photons.

Our analysis was performed in the regime where the motion can be treated semiclassically and is valid in the Doppler-cooling regime. In the future our theoretical studies could be extended to the sub-Doppler regime, where the calorimetry signal may become even more sensitive to temperature changes, e.g., when the ion crystal is undergoing polarization-gradient cooling\,\cite{PolGradCooling_Wineland_1992,PolGradCooling_Roos_2020,PolGradCooling_Li_2021}. Under these conditions, we conjecture that the quantum fluctuations can no longer be neglected. Moreover, we expect that one will include the modification of the photon emission due to collective effects in the ordered array of ion emitters \cite {WOLF2020,RICHTER2021}.

We finally remark that the intensity of the scattered light allows one to solely determine the mean energy, but gives no information on whether the crystal is in thermal equilibrium or whether the crystal eigenmodes are thermalized at the same temperature. Future work should investigate how the latter information could be extracted by measuring the light emitted from different ions within the crystal,  employing detectors which selectively collect the light from individual emitters within the crystal. Alternatively, this information could be extracted from impurity ions embedded in the crystal \cite{MorigiWalther2001,Monroe_mixedSpecies}.

\section*{Acknowledgements}

We thank the TACTICa collaboration members for useful discussions. This work was funded by the Deutsche Forschungsgemeinschaft (DFG, German Research Foundation), Project No. 429529648 TRR 306 QuCoLiMa (Quantum Cooperativity of Light and Matter). This work was also partly supported by the Cluster of Excellence Precision Physics, Fundamental Interactions, and Structure of Matter (PRISMA+ Grant No. EXC 2118/1) funded by the DFG within the German Excellence Strategy (Project No. 39083149), by the DFG Reinhart Koselleck project, and through the DIP program (Grants No. SCHM 1049/7-1 and No. FO 703/2-1).  W.H. acknowledges support from a research fellowship from the DFG, (Grant No. HA 8894/1-1). W.L. acknowledges financial support from the China-Germany Postdoctoral Exchange Program (Grant No. 2018018).

\section*{Appendix}

\appendix

\section{Estimate for the Coulomb logarithm} \label{app_cl}

The factor $\Lambda_\text{C}$ in Eq. \eqref{eqn_dedx} can be approximated~\cite{POTH,ZWICKNAGEL} by \mbox{$\Lambda_\text{C}\approx\int_{b_\text{min}}^{b_\text{max}} \frac{db}{b}$}, where $b$ is the scattering impact parameter and $b_\text{max}$ is usually estimated by the screening length of the Coulomb potential $b_\text{max}\approx v_\text{i}/\nu_\text{p}$ and $b_\text{min}$ by the maximal momentum transfer $b_\text{min}=e^2/4\pi\epsilon_0M_\text{Th}v_\text{i}^2$. Substituting the parameters of the TACTICa experiment and $v_\text{i}=v_{\text{i}2}$, we obtain $\Lambda_\text{C}\approx\ln\left(\frac{b_\text{max}}{b_\text{min}}\right)\approx 9$.

\section{Normal modes of the ion crystal} \label{app_osc}

The range of the normal-mode frequencies $\omega_m$ can be estimated as follows (see also Refs. \cite{James_1998,Rohde_2001}). The highest-frequency modes usually correspond to the ions oscillating in antiphase with respect to their nearest neighbors. This implies that, for the highest-frequency modes, $\omega_m$ can be estimated by $\sqrt{2}\omega_h$, where $\omega_h$ is the frequency of the quasiharmonic potential that each ion experiences in the ion crystal. This potential is mainly governed by the Coulomb repulsion from the nearest neighbors, hence the frequency is $\omega_h=\sqrt{e^2/\pi\epsilon_0Ml^3}$, where $l$ is the average distance between the ions and $M$ is the mass of a crystal ion. For the lowest-frequency modes, $\omega_m$ can be approximated by the external harmonic trapping potential with frequency $\omega_\text{a}$ in the axial direction and $\omega_\text{r}$ in the radial direction.

The difference of the trapping frequencies in the axial and radial directions usually leads to cigar-shaped ion crystals. We assume that due to this asymmetry, the normal modes of the crystal split into two groups having a preferred direction of oscillation in either the axial or the radial direction \cite{MorigiFishman:2004}. In each direction, we assume the dispersion relation to be linear similar to the Debye model\,\cite{ashcroft} such that the mode frequencies can acquire the values $\omega_{\min}+n(\omega_{\max}-\omega_{\min})/(N_\text{Ca}-1)$, where $n$ is an integer $n\in[0,N_\text{Ca}-1]$ and $\omega_{\min}$ ($\omega_{\max}$) is the minimal (maximal) mode frequency in the given direction.

Using \mbox{$l\approx4\mu$m}, $M=M_\text{Ca}$, \mbox{$\omega_\text{a}\sim 2\pi\times100$\,kHz} and $\omega_\text{r}\sim 2\pi\times1$\,MHz, we obtain $\omega_h\sim 2\pi\times15$\,MHz and, hence, for the approximate range of normal-mode frequencies $\omega_m/(2\pi)\!\in$ [100\,kHz,20\,MHz] for the axial direction and $\omega_m/(2\pi)\!\in$ [1\,MHz, 20\,MHz] for the radial direction, respectively. 

Let us also remark that micromotion\,\cite{micromotion}, which is expected to be strong in three-dimensional ion crystals, where the equilibrium positions of ions can be far from the trap axis, is included in our estimates of temperatures achievable with Doppler cooling~\cite{Champenois_MassSpectroscopy}.

\section{Timescale for Doppler cooling} \label{app_ld}

\begin{figure}[t]
    \centering
    \includegraphics[width=0.8\columnwidth]{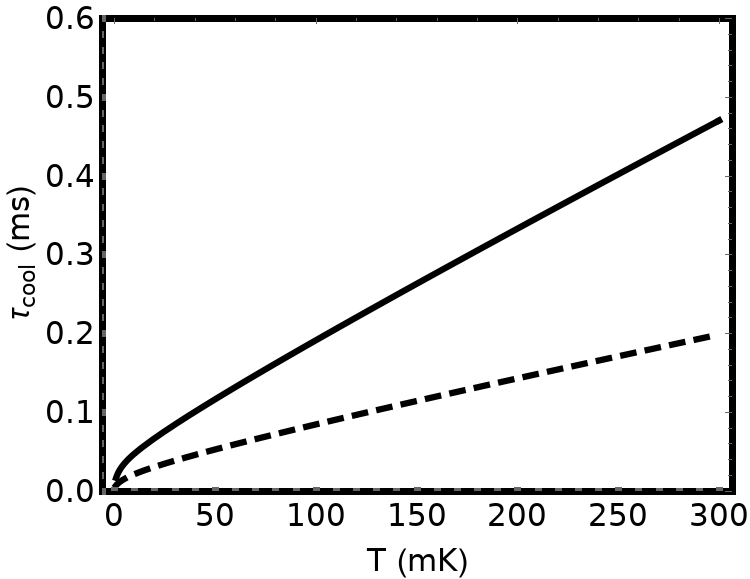}
    \caption{Rate of change of the motional energy per ion [cf. Eq. \eqref{eq_tau_cool}] as a function of the initial temperature and for the parameters of the TACTICa experiment and \gm{$\Delta = 3\gamma/4$}. The saturation is $s=0.25$ \gm{(solid line)} and \gm{$s=1$ \gm{(dashed line)}}.}
    \label{fig:tau}
\end{figure}

In the semiclassical model the cooling of the ion crystal is described with the radiation pressure force $\textbf{F}$ exerted on the individual ions, such that their momentum \gm{evolves} as
\begin{equation}
    \frac{d\textbf{p}}{dt} = \textbf{F} = \hbar \textbf{k} \gamma \rho_{ee}. 
\end{equation}
\gm{where $\gamma\rho_{ee}$ is the rate of photon scattering and $\hbar {\bf k}$ is the average change of linear momentum per scattering event \cite{stenholm1986semiclassical}.}
This yields a rate of change of the motional energy $E=\textbf{p}^2/2m$ of the ion crystal as
\begin{equation}
    \tau_\text{cool}^{\gm{-1}} = \frac{dE}{dt}/E = 2 \sqrt{\frac{\hbar \omega_R}{E}} \gamma \rho_{ee}. \label{eq_tau_cool}
\end{equation}
For the parameters of the TACTICa experiment and temperatures on the order of $100$~mK, this rate of change is on the order of \gm{$200 \mu$s} for a saturation parameter of $s=0.25$ (see Fig. \ref{fig:tau}). 

\bibliography{thorium}

\end{document}